\documentclass[5pt]{article}

\usepackage[letterpaper]{geometry}
\usepackage{spconf,amsmath,epsfig}
\usepackage{enumitem}
\usepackage[colorlinks,linkcolor=blue,anchorcolor=blue,citecolor=blue]{hyperref}

\usepackage{fancyhdr}
\begin{document}\sloppy
	
% Example definitions.
% --------------------
\def\x{{\mathbf x}}
\def\L{{\cal L}}

% Title.
% ------
\title{%SenseCare: An Artificial Intelligence Platform for Clinical Researches towards Smart Health
    SenseCare: A Research Platform for Medical Image Informatics and Interactive 3D Visualization}
%
% Single address.
% ---------------
%\name{Qi Duan, Guotai Wang, Rui Wang, Chao Fu, Xinjun Li, Maoliang Gong, Xinglong Liu, Qing Xia, 
%Xiaodi Huang, Zhiqiang Hu, Ning Huang and Shaoting Zhang}
\name{%
    \begin{tabular}{@{}c@{}}
        Qi Duan, Guotai Wang, Rui Wang, Chao Fu, Xinjun Li, Na Wang, Yechong Huang, Xiaodi Huang, \\  
 Tao Song, Liang Zhao, Xinglong Liu, Qing Xia, Zhiqiang Hu, Yinan Chen and Shaoting Zhang
\end{tabular}}
 \address{SenseTime Research\\
    smarthealth-contact@sensetime.com}
%
% For example:
% ------------
%\address{School\\
%	Department\\
%	Address\\
%   Email}
%
% Two addresses (uncomment and modify for two-address case).
% ----------------------------------------------------------
%\twoauthors
%  {A. Author-one, B. Author-two\sthanks{Thanks to XYZ agency for funding.}}
%	{School A-B\\
%	Department A-B\\
%	Address A-B}
%  {C. Author-three, D. Author-four\sthanks{The fourth author performed the work
%	while at ...}}
%	{School C-D\\
%	Department C-D\\
%	Address C-D\\
%   Email}
%

\maketitle

% --------------------------------------------------------------------------

\begin{abstract}
    Clinical research on smart health has an increasing demand for intelligent and clinic-oriented medical image computing algorithms and platforms that support various applications. To this end, we have developed SenseCare research platform, which is designed to facilitate translational research on intelligent diagnosis and treatment planning in various clinical scenarios. To enable clinical research with Artificial Intelligence (AI), SenseCare provides a range of AI toolkits for different tasks, including image segmentation, registration, lesion and landmark detection from various image modalities ranging from radiology to pathology. In addition, SenseCare is clinic-oriented and supports a wide range of clinical applications such as diagnosis and surgical planning for lung cancer, pelvic tumor, coronary artery disease, etc. SenseCare provides several appealing functions and features such as advanced 3D visualization, concurrent and efficient web-based access, fast data synchronization and high data security, multi-center deployment, support for collaborative research, etc.  In this report, we present an overview of SenseCare as an efficient platform providing comprehensive toolkits and high extensibility for intelligent image analysis and clinical research in different application scenarios. We also summarize the research outcome through the collaboration with multiple hospitals.   
\end{abstract}
	

	% --------------------------------------------------------------------------
	
\section{Introduction}
\label{sec:intro}
	%Clinical research has a high value to society, as the medical and healthcare services we can enjoy is built on the efforts of numerous physicians, scientists and other medical professionals. They have been tirelessly investigating disease trends and risk factors, experimenting with potential treatment planning, and thus contributing to future policy making. However, physicians confront a variety of challenges with regard to their involvement in clinical research. Even physicians may overcome the problems of insufficient financial support, inadequate clinical cases, or complicated compliance processes, busy patient practices leave them with limited time for research. 
	With the development of medical imaging techniques and computer science, computer-aided systems for medical image analysis and downstream diagnosis and treatment decision have been playing an increasing role in clinic practices. In recent years, Artificial Intelligence (AI) has lead to a revolution of image analysis and pattern recognition, and has a huge potential to be applied for more efficient and intelligent medical image computing in a wide range of medical departments towards smart healthcare. However, before AI is used in clinic practice, extensive  research studies are needed through the collaboration between clinicians, radiologists, pathologists, surgeons, AI scientists and engineers, which can validate the effectiveness, robustness, reliability and security of AI systems. To this end, a research platform that supports different medical image processing tasks and intelligent medical image computing  algorithms for various clinic applications are highly desirable.
	\\
	\\Despite the availability of several existing platforms for developing AI algorithms for general image recognition or medical image computing, they are not clinic-oriented and have limited support for clinical research.  For example, TensorFlow~\cite{Abadi2016a}, Pytorch~\cite{Paszke2019}, TensorLayer~\cite{Dong2017} and Keras~\cite{Chollet2015} are general deep learning libraries that provide low-level functions to develop complex deep learning models without specific functionality for medical image computing. Some other libraries such as NiftyNet~\cite{Gibson2018} and DLTK~\cite{Pawlowski2017} are developed for deep learning with medical images, but they are mainly designed for AI algorithm developers and there is no graphic user interface, which is difficult for clinicians and radiologists to use in specific clinic applications. On the other hand, several medial image analysis solutions have also been developed in the past decades.  Tools that solve a specific medical image processing task such as segmentation (e.g., NiftySeg~\cite{Cardoso2012}), registration (e.g., NfityReg~\cite{Modat2010b} and elastix~\cite{Klein2010a}) and visualization (e.g., VTK~\cite{Schroeder2005a}) can be used for a part of a clinic application pipeline, but still not ready-to-use for clinic researchers. Some research platforms such as MITK~\cite{Nolden2013}, NifTK~\cite{Clarkson2015} and 3D Slicer~\cite{Pieper2004} provide several traditional medical image analysis tools and 3D visualization for image guided intervention. However, these platforms have almost no AI models and are not suitable for clinical research on intelligent medical image computing towards smart healthcare.  In the era of big data, a desired clinical research platform based on AI does not only need to provide AI models, but also require data exchange with other image management systems, functionality across diverse image modalities, availability of sophisticated 3D visualization and high extensibility and portability for different clinic scenarios. 
	\begin{figure*}[t]
		\centering
		\includegraphics[width=1.0\linewidth]{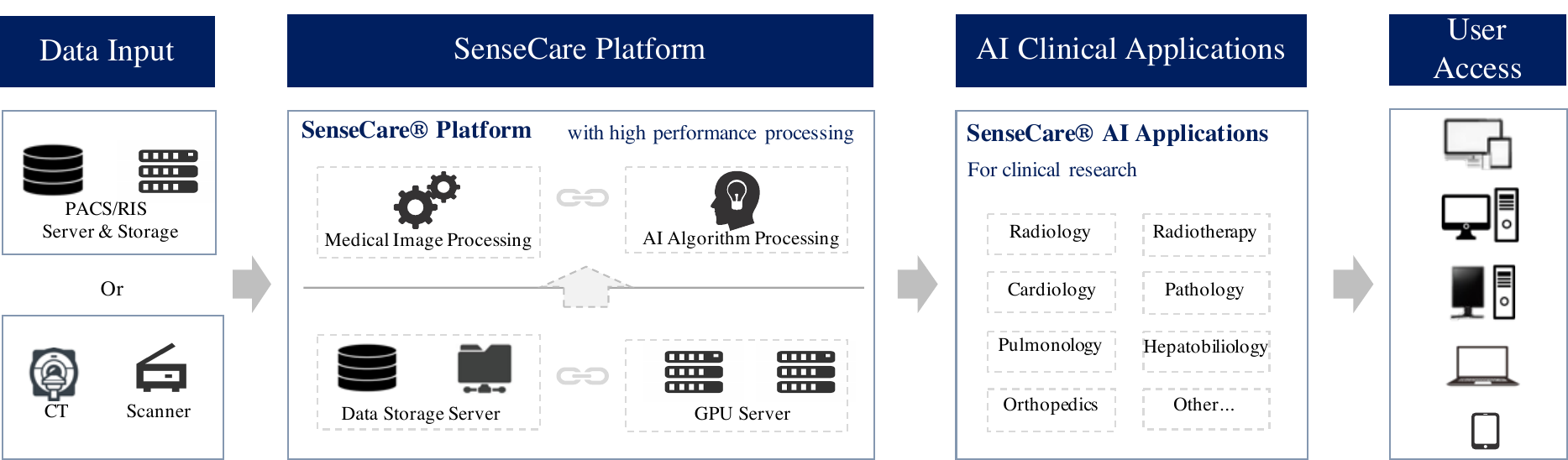}
		\caption{Architecture of SenseCare platform. }
		\label{fig:architecture}
	\end{figure*}
	\\
	\\To solve these problems, we introduce the
	SenseCare Smart Health Platform (SenseCare for short) that aims to take advantages of state-of-the-art AI techniques to foster researchers from different clinical departments to implement innovations for improvement through the whole process of clinical diagnosis, treatment planning and rehabilitation management. SenseCare has the following appearing functionalities for clinical research towards smart healthcare:
	\begin{enumerate}
		\item \textbf{Equipped with powerful AI toolkits.} SenseCare has integrated a wide range of ready-to-use AI models for different medical image computing tasks such as image segmentation, registration, lesion and landmark detection. It also provides convenient tools for image annotation, model training and validation for developing deep learning algorithms customized by users. 
		\item \textbf{Support for various applications in clinical research.} SenseCare is clinic-oriented and can be easily used by clinicians, radiologists and other clinical researchers with several functional modules such as advanced 3D visualization, data and research task management. Its use is not constrained to a specific department or laboratory. SenseCare can be adapted to various clinic applications including lung cancer diagnosis and treatment planning, pathological image analysis, coronary artery disease diagnosis and modeling, surgical planning and simulation for pelvic tumors, etc. 
		\item \textbf{Easy to access with high concurrency}. Services of SenseCare can be accessed on a browser and it does not rely on a specific device or operating system. Its high concurrency allows hundreds of users to use the intelligent image computing services simultaneously. 
		Its multi-center deployment makes it much easier for researchers from multiple departments in different locations to collaborate and perform multi-center studies. 
		\item \textbf{Support for various data modalities with high synchronization and security.} SenseCare can deal with various data modalities covering radiological and  pathological images, medical records and other types of data. It also provides efficient synchronization of data from different information systems and across multiple centers, and uses encryption algorithms to protect user accounts and data from latent risk factors. 
	\end{enumerate}
	In this paper, we introduce favorable features distinguishing SenseCare  from other platforms and workstations targeting at AI-driven clinical research on intelligent diagnosis and treatment towards smart healthcare. SenseCare has increasingly assisted to achieve outputs in several research projects including quantitative analysis of cardiac function~\cite{Li2019c}, assessment of knee
	articular cartilages~\cite{Tan2019}, pathological image analysis~\cite{Li2019a, Li2019,  Qu2019}, lung cancer diagnosis~\cite{Yu2019a, Zhang2019},  quantitative brain tumor assessment~\cite{Wang2019a,Wang2019b}, spine image analysis~\cite{Gao2019,Chen2020} and radiotherapy planning for head and neck cancer~\cite{Gao2019a,Lei2019}, etc.
	\\
	\\This paper is organized as follows: In Section~\ref{sec:acrhitecture}, we give a brief summary of the architecture of SenseCare, which is followed by detailed descriptions of basic functional modules in Section~\ref{sec:functional modules}.  We then introduce SenseCare's AI toolkits in Section~\ref{sec:AI_algirithms}. In Section~\ref{sec:clinical_app}, we show several examples of clinical applications powered by SenseCare. Finally, conclusions are given in Section~\ref{sec:conclusion}. 
	
	\section{Architecture Of SenseCare}\label{sec:acrhitecture}
	As shown in Fig.\ref{fig:architecture}, SenseCare provides a wide range of artificial intelligence algorithms based on deep learning for learning from and analyze different kinds of medical big data. With these AI algorithms and some other supporting modules such as data management, users can easily adapt the deep learning framework to various clinical applications efficiently. SenseCare also provides advanced visualization of medical images that enables users to analyze complex anatomies and segmented structures. These  modules are combined with a browser/server architecture and multi-center deployment so that they are accessible on different kinds of devices and at various locations. 
	\\
	\\
	The architecture of SenseCare follows a modular structure that consists of
	% Although original data are strictly restrained within the hospital, iPads, smart phones, laptops and other portable devices allow users to access SenseCare and retrieve, review, or manipulate on it. 
	%In pursuit of better interpretations and understandings of multimodal medical images, SenseCare is designed to support versatile visualizations and provide advanced processing capabilities, meanwhile its architecture follows a modular and layered approach to be capable of implementing new algorithms and integrating specific useful functions. 
	%The well-structured architecture of SenseCare is comprised of 
	three layers: 1) basic functional modules such as data management and visualization, web-based access and multi-center deployment, 2) advanced AI toolkits that include libraries for model training and many built-in artificial intelligence algorithms for image segmentation, registration, lesion detection, etc., and 3) application scenarios that adapt the basic functional modules and AI algorithms to deal with different clinical tasks such as computer assisted diagnosis of the lung and surgical planning for bone tumors.

	\section{Basic Functional Modules}\label{sec:functional modules}
	\subsection{Data Support}
	
	%This section will focus on the lower level of the SenseCare, which consists of four basic functional modules: data, visualization, network and framework. 
	%When it comes to data module, SenseCare supports multi-modality imaging data, establishes data synchronization and guarantees strict data security. Visualization module manages and presents data in a visual form to facilitate image analysis and interpretation. Network connection ensures user-end accessibility and portability to SenseCare. Different from other research or clinical platform, SenseCare is embedded with an innovative deep learning framework SenseParrots developed by SenseTime, which makes model training easier and more efficient. 
	Due to different imaging mechanisms, medical images are acquired in a variety of modalities or protocols. Being able to deal with these images is important for an intelligent image computing system to be useful in a clinical research environment. Besides images, other medical data such as medical records are also important for diagnosis of diseases and clinical follow-up. SenseCare supports medical images in different modalities, and provides efficient and convenient synchronization of data from different information systems in the hospital or medical center.
	
	\subsubsection{Support of Various Imaging Modalities}
	SenseCare supports intelligent analysis of in images in various modalities ranging from radiological images to pathological images.  For radiological images, it allows efficient import, query, retrieval, and storage of clinical images using DICOM protocols and structures. Major radiological images including Computed Tomography (CT), Magnetic Resonance Imaging (MRI), Digital Radiology (DR) and Positron Emission Tomography (PET) are all well supported by SenseCare in different clinical applications.
	\\
	\\
	For pathological images, SenseCare supports several image formats including SVS, TIFF, VMS, NPDI, KFB and others. It also provides a series of functions of import, query, retrieval, storage, management, common measurements and analysis to help pathologists perform diagnosis in a more efficient and intuitive way. 
	
	\subsubsection{Data Synchronization and Security}
	Medical data are commonly stored in different systems such as the Picture Archiving and Communication System (PACS) and Radiology Information System (RIS). Synchronization of data between these systems and the image computing workstation is critical in a clinical research environment. 
	To facilitate more efficient and functional workflow for clinical researchers and medical practitioners, 
	SenseCare provides users with improved access by efficient data synchronization. It is capable of synchronizing data from PACS, RIS and other common information systems in hospitals without disturbing the original clinical workflows. When newly acquired data are transmitted or changes in status take place in these information systems, SenseCare will synchronize the updated information and present the users with the latest information automatically. 
	\\
	\\Besides real-time data synchronization, SenseCare also supports pulling data directly from PACS/RIS based on user-defined rules. For example, users can designate image modality and type or time range, and send queries to fetch the data they want from a database. They can also synchronize all the follow-up data after fetching the data of a patient based on patient ID. As for pathological applications,  SenseCare can be seamlessly connected to common pathological scanners to obtain data, and  users can retrieve  digital slides by giving their labels. 
	\\
	\\In addition, encryption algorithms embedded in SenseCare reliably manage and protect user accounts and data from potential risk factors. Although data and images are utilized and transmitted over a shared database on which multiple users manipulate, SenseCare strictly ensures data security, integrity and consistency. 
	
	\subsection{Advanced Visualization}
	\begin{figure}
		\centering
		\includegraphics[width=1.0\linewidth]{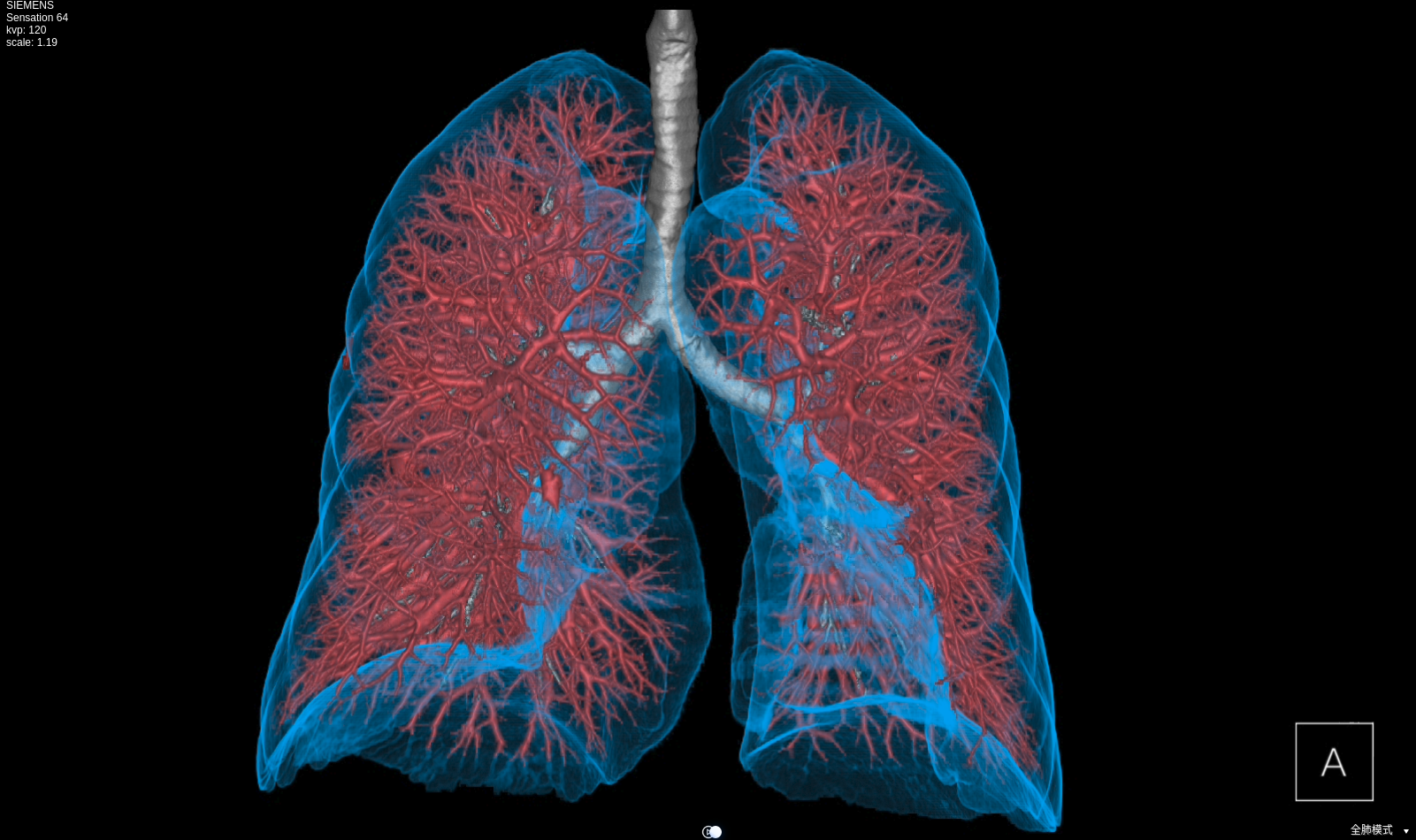}
		%\centerline{\epsfig{figure=pictures/3dlung.png,width=8.5cm}}
		\label{fig:lung_reconstruction}
		\caption{3D reconstruction of the lung in SenseCare.}
	\end{figure}
	%SenseCare is well integrated with visualization capabilities and therefore provides researchers and clinicians with an easy visualization tool to learn and use. By synchronizing and displaying medical images, SenseCare presents data and information in a more explicit way thus improving the information interpretation. With high resolution visualization, users can glean detailed information and utilize it more efficiently. %For example, display of lung CT images will help clinicians and radiologists locate nodules easily. %Actually, besides manual annotation, SenseCare are capable of detecting lesions and providing further quantitative information. Those image analysis algorithms are the key factors that make SenseCare not only an imaging reader but also a comprehensive analysis workstation. 
	SenseCare provides advanced 3D reconstruction and visualization of medical images to facilitate the analysis of complex anatomies and segmented structures, which presents data and information in a more explicit way thus improving the information interpretation.  Comprehensive methods such as Maximum Intensity Projection (MIP), Minimum Intensity Projection (MinIP), Multi-Planar Reconstruction (MPR), Curved Planar Reformation (CPR) and 3D volume rendering are available for users to perform 3D visualization and enhance interactivity. These capabilities play an important role in clinical diagnosis meanwhile contributing to surgical planning, simulation and navigation, as well as radiotherapy planning, etc. 
	\\
	\\For instance, by providing 3D visualization of thoracic anatomies and intervention path recommendation, SenseCare assists clinicians in performing preoperative planning in a more intuitive and convenient way. Fig.~\ref{fig:lung_reconstruction} shows an example of 3D visualization of lung  structures performed by SenseCare. The system provides automatic rotation of 3D models so that users can observe the models from different views for a comprehensive understanding.  Users can also interactively zoom, pan and rotate the view and click to hide tracts or vessels in real time. 
	
	\subsection{Concurrent and Efficient Web-based Access}
	
	SenseCare is designed with high performance concurrency. The network communication structure of SenseCare enables hundreds of users to concurrently perform high-resolution image rendering and 3D post-progressing thus satisfying multiple usage needs. With servers deployed, SenseCare platform allows over 1,000 concurrent users to review and retrieve medical images while its comprehensive toolkits for advanced image post-processing are designed to support more than 160 users simultaneously. %, which actually outperform prestigious medical imaging institutions and universities worldwide. 
	\\
	\\SenseCare can be used in different devices and operating systems and no plugins are needed. By adopting a browser/server architecture, it offers a truly seamless user experience and eliminates the need for multiple logons. Users can access not only MPR, MIP/MinIP, CPR tools, but also the full range of three-dimensional capabilities through HTML5 websites. SenseCare grants radiologists and physicians with efficient access and workflow-boosting benefits even when working from iPad, smartphone or laptop. 
	
	\subsection{Multi-center Deployment}
	Collecting data from multiple centers is important for developing robust algorithms and large-scale clinic studies. To support such research that requires a collaboration between researchers from different locations, SenseCare can be deployed at multiple centers. This favorable feature distinguishes  SenseCare from traditional image computing workstations that are located at a single institution. 
	\\
	\\
	In addition, SenseCare's multi-center deployment facilitates the data collection and access process. Various types of data from different centers can be collected and cleaned under standardized rules. Data scientists, clinical researchers, and pharmacists are all able to participate in the process while respecting the original clinical process. 
	%\\In the age of information and big data, polycentric platform helps clinical research regarding innovative medicine, medical equipment, surgical plans, and so on. It becomes much easier for researchers
	\subsection{Support for Collaborative Research}
	To facilitate the collaboration between researchers, SenseCare provides a specific Document Management System (DMS) to make it easier to organize, secure, capture, digitize, tag, approve and complete tasks with research-related files. It can handle a large amount of papers and images and cope with different workflows by supporting advanced user permission management and task management. This efficient tool enables researchers to manage multiple sophisticated research projects easily, and helps a ground of researchers collaborate with each other and exchange knowledge for efficient accomplishment of comprehensive research projects. 
	\\
	\\
	For user permission management, SenseCare supports different levels of users. Admin users can create normal users and set permissions for other normal users to access, view, manipulate, share and remove folders and files. Nonetheless, Admin users can change the permission if needed. 
	For task management, SenseCare's DMS allows admin users to break down tasks and delegate tasks to normal users. It also allows users to track task progress from the beginning to the end and set small milestones to make sure the whole project will be finished on time. Users can prioritize, organize and set deadlines for themselves and are able to draw together the resources they need to achieve their research  goals. 
	
	%\\
	%\\Instead of writing hundreds of lines of codes,  users are capable of using SenseParrots deep learning framework. With a collection of pre-built and optimized components, it is easier for users and developers to build and train models. Besides, SenseParrots deep learning framework helps protect intellectual property rights. Users can efficiently generate, publish and utilize their own libraries on the platform. 
	%\\
	%\\A typical workflow for clinical research follows data synchronization, annotation, training, application. With SenseParrots and based on their libraries, users are capable of performing semi-automatic annotation and design, build, and train new deep learning models interactively and to eventually evaluate them by feeding data to them. 
	
	%\\On the basis of SenseParrots, SenseCare supports a range of state-of-the-art artificial intelligence algorithms for medical image computing tasks, including image segmentation, registration, landmark and lesion detection, etc. Details of these algorithms and their applications will be described in Section~\ref{lab:research_platform} and Section~\ref{lab:clinical_app}. 
	
	\section{Artificial Intelligence Toolkits}\label{sec:AI_algirithms}
	Recent years have seen a fast growing of novel deep learning algorithms for medical image computing tasks~\cite{Shen2017}, which play an important role for more accurate and efficient diagnosis and treatment planning and assessment. To boost research on deep learning algorithms for medical imaging, SenseCare has integrated commonly used traditional image processing algorithms with advanced state-of-the-art deep learning models. Based on these features, SenseCare can largely improve the efficiency of development and validation of novel deep learning algorithms by users. In this section, we introduce SenseCare's AI toolkits for medical image computing, including tools for users to develop new algorithms, and built-in deep learning models for image segmentation, registration and lesion detection, etc. 
	
	\subsection{Deep Learning Framework}
	SenseCare's AI toolkits are constructed on the basis of SenseParrots, a deep learning framework independently developed by SenseTime. While matching mainstream frameworks such as TensorFlow~\cite{Abadi2016a} and PyTorch~\cite{Paszke2019}, it also has advantages in ultra-deep networks and ultra-large-scale data model training. The integrated optimization from the underlying layer to the SenseParrots framework enables the platform to outperform others under the same level of configuration. Currently, the testing of a single training task uses more than 1,000 GPU card for parallel training. It takes less than 1.5 minutes to complete the AlexNet neural network training, surpassing the previous fastest 4-minute record~\cite{Sun2019}. 
	
	\subsection{Annotation Tools for Model Training}
\begin{figure}
		\begin{minipage}[b]{1.0\linewidth}
			\centering
			\centerline{\epsfig{figure=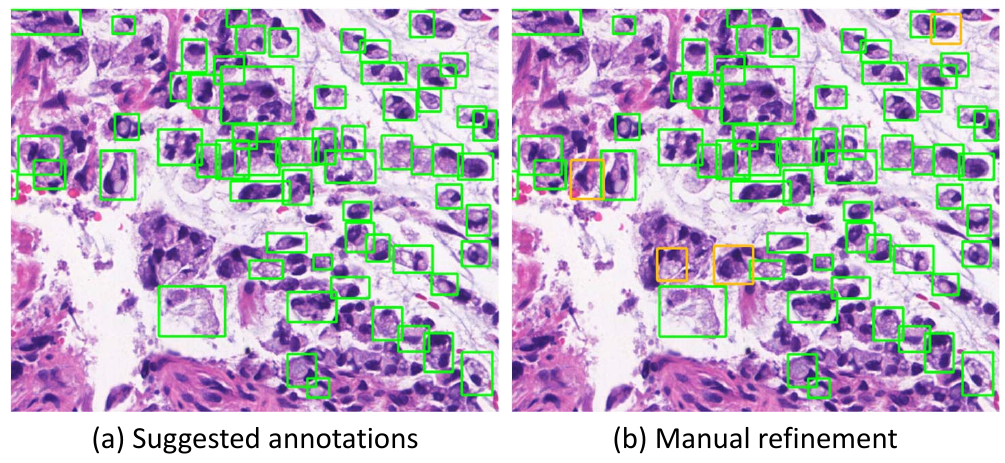,width=\linewidth}}
		\end{minipage}
		\caption{Annotation of signet ring cell carcinoma.}
		\label{fig:annotation_ring}
	\end{figure}

	Annotations of a large set of training images are critical for achieving high performance of deep learning models. To facilitate the development of these models, SenseCare provides a set of off-the-shelf tools for efficient image annotation, such as contouring tumors and organs for segmentation tasks, bounding box annotation for object and lesion detection tasks. 
	\\
	\\
	The annotation tools in SenseCare are available for both radiological images and pathological images and they provide a variety of annotation types. For example, users can choose from different interactive styles including mouse click points, rectangles, circles, ellipses, polygons or hand-drawn shapes based on the characteristics of the target.
	Since manual annotation is time-consuming and annotation may vary from different annotators' inputs, SenseCare also supports semi-automatic annotation, which combines the efficiency of automatic annotation and the accuracy of manual annotation. Annotators can start from automatically generated annotations and only need to provide few interactions to obtain refined annotations~\cite{Luo2021mideepseg,Wang2020ugir}, which can largely reduce  burdens of annotators and improve the efficiency. As an example,  Fig.~\ref{fig:annotation_ring} shows  efficient annotation of signet ring cell carcinoma in SenseCare, where algorithms suggested some annotations in Fig.~\ref{fig:annotation_ring}~(a), and the annotator only provided manual refinement to obtain accurate annotations for a slide in Fig.~\ref{fig:annotation_ring}~(b).

	%\begin{figure}
	%	\begin{minipage}[b]{1.0\linewidth}
	%		\centering
	%		\centerline{\epsfig{figure=pictures/nodules.png,width=\linewidth}}
	%	\end{minipage}
	%	\caption{Annotation of pulmonary nodules.}
	%	\label{fig:annotation_nodule}
	%\end{figure}
	
	\subsection{Detection}
	\begin{figure}[htbp]
		\begin{minipage}[b]{1.0\linewidth}
			\centering
			\centerline{\epsfig{figure=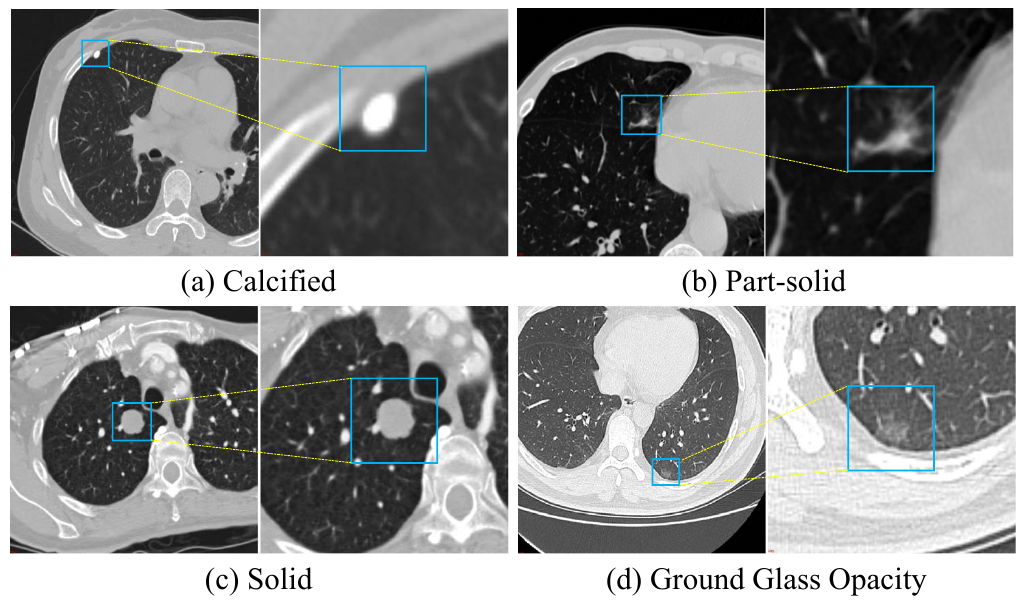,width=\linewidth}}
		\end{minipage}
		\caption{Lung nodule detection by SenseCare.}
		\label{fig:lung_nodules}
	\end{figure}
	Automatic detection of objects of interest is a common task for computer assisted diagnosis systems. SenseCare has several built-in deep learning models for object detection tasks, such as landmark and lesion detection in 3D radiological images and cancer cell detection in histopathological images. 
	\\
	\\For example, automatic localization of vertebrae in CT is important for image-guided diagnosis, pre-operative planning and post-operative evaluation. Deep learning models such as 3D Fully Convolutional Neural Networks (FCN) are  embedded in SenseCare for accurate vertebrae localization~\cite{Chen2020}. The model also takes advantages of prior knowledge such as spatial and sequential constraints to obtain high robustness in challenging cases. In addition, SenseCare is able to predict tumor invasiveness and malignant in Ground Glass Opacity (GGO) on the basis of its lung nodule detection model~\cite{Yu2019a}. Fig.~\ref{fig:lung_nodules} shows an example of lung nodule detection by deep learning models in SenseCare.
	For signet ring cell carcinoma detection from Hematoxylin and Eosin (H\&E) stained Whole Slide Images (WSI), SenceCare is integrated with bottom-up approaches~\cite{Li2019a} that obtain cell instance masks first and then derive bounding boxes for each instance, which is more accurate than the general RCNN-based detection methods.

	\subsection{Segmentation}
	
	Image segmentation is essential for most clinical applications such as accurate modeling of anatomical structures, quantitative measurement of tumor volumes, planning of radiotherapy and surgical treatment.  Its output has a large impact on the downstream workflows. However, due to the low contrast between the target tissue and its surroundings, inhomogeneous appearance, complex shape variation and image noise, accurate segmentation is extremely challenging and  traditional image segmentation algorithms are often faced with large regions of over- and under-segmentation.
	\\
	\\
	Supported by its deep learning-based image segmentation models, SenseCare can overcome these challenges and has obtained state-of-the-art performance in a range of segmentation tasks.  For example, to segment the complex structures of pulmonary vessels from CT images, SenseCare is equipped with a multi-view-based 2.5D network with a low complexity, which outperformed other contemporary networks by a large margin on the LIDC dataset~\cite{Cui2019a}. To segment intervertebral discs from MR images, a novel multi-resolution path network with deep supervision is included in SenseCare and it achieved superior performance on the MICCAI 2018 IVDM3Seg challenge dataset~\cite{Gao2019}. Several specific CNN models are also developed for other applications, such as multiple Organs-at-Risk (OAR) segmentation from CT for radiotherapy planning~\cite{Gao2019a}, segmentation of optic disc and cup for glaucoma diagnosis~\cite{Liu2019}, nuclei instance segmentation in histopathological images~\cite{Li2019, Qu2019} and
	cartilage segmentation from MR images for osteoarthritis assessment~\cite{Tan2019}. These different models are ready-to-use and serve as strong baselines for the above specific applications, and they can be easily adapted to new segmentation tasks based on the user's research interests.

	\subsection{Image Registration}
	In clinical applications, images acquired in different modalities are often need to be fused to provide sufficient information for diagnosis and treatment decision. In addition, a patient may be scanned several times at different stages of a disease to obtain a better understanding of the evolution of the disease.  Therefore, it is necessary to register two or more images into a common spatial coordinate system for better interpretation of anatomical and pathological characteristics thus improving diagnosis and treatment for patients.
	\begin{figure}
		\begin{minipage}[b]{1.0\linewidth}
			\centering
			\centerline{\epsfig{figure=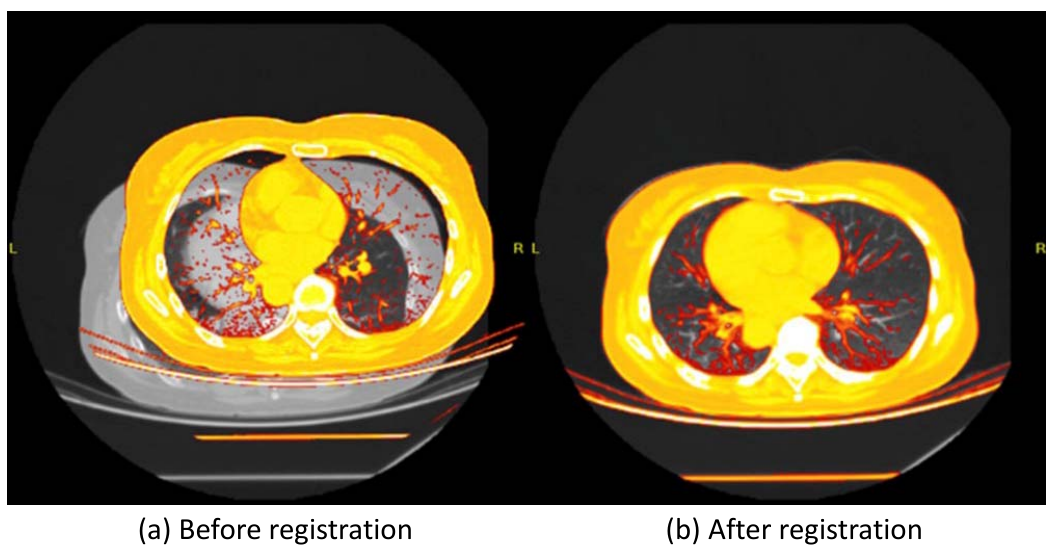,width=\linewidth}}
		\end{minipage}
		\caption{An example of non-rigid  lung image registration in SenseCare.}
		\label{fig:registration}
	\end{figure}
	\\
	\\
	SenseCare provides several registration algorithms for different tasks.  For example, the combination of MR and CT is quite useful since the former is better suited for delineation of tumor regions while the latter is needed for accurate computation of the radiation dose.  Registration between MR and CT images supported by SenseCare makes it more efficient and accurate to obtain radiotherapy planning. Registration algorithms embedded in SenseCare include both rigid and non-rigid registration between images from a single modality or multiple modalities. 
	Rigid methods are useful for the the registration in the presence of rigid bodies such as bones. Non-rigid registration is used for applications such as correcting soft-tissue deformation during imaging and modeling the dynamic heart. Fig.~\ref{fig:registration} shows an example of non-rigid lung image registration in SenseCare.
	%\\
	%\\In most cases, the visualization and analysis of human tissues and organs are based on a sequence of slices obtained with high resolution using CT, MRI or other modalities. However, due to the capability of the image devices and other limitations, the space between consecutive slices is usually larger than the pixel size.  For better application of such data for 3D reconstruction, establishing some kind of interpolation between the slices is rather crucial in image processing. A great variety of interpolation methods haven been put forward and SenseCare is embedded state-of-the-art algorithms to improve interpolation results thus facilitating the image processing. Adaptive algorithm basically uses some of the image features and improves interpolation results.

	\subsection{Docker Integration}
	Users of SenseCare can employ the built-in deep learning models mentioned above for several image computing tasks, and they can also develop their own models with the help of SenseCare's training and testing pipelines. When the user creates a new algorithm,  SenseCare provides a dockerized version of the algorithm, and can package it with all the dependencies together in the form of containers, so that the algorithm can work seamlessly in any new environment. Therefore, researchers can focus on the development of AI algorithms without worrying about the testing and production environment. 
	Since dockers are lightweight, SenseCare makes it more convenient and efficient for users to develop, test and deploy algorithms and deep learning models for various clinical applications. %They use the resources that are available and have higher performance without waste of computing resources. This tool is sure to be handy when users want to utilize promoted algorithms based on their semi-automatic annotation. 
	
\section{Medical Image Analysis Challenges supported by SenseCare}~\label{sec:miccai_challenges}
    Supported by the above basic functional modules and AI toolkits with high extensibility in SenseCare, we successfully organized two medical image analysis challenges in conjunction with International Conference on Medical Image Computing \& Computer Assisted Intervention (MICCAI) 2019: DigestPath and StructSeg, which facilitated researches on  pathological image analysis and 3D medical image segmentation for diagnosis and radiotherapy planning of cancers, respectively.  

\subsection{The DigestPath Challenge}
  The DigestPath 2019 Challenge\footnote{https://digestpath2019.grand-challenge.org/Home/} consists of two tasks for  digestive-system pathological cell detection and tissue segmentation. The first task is signet ring cell detection, and a total of 155 patients' WSIs with size of 2000$\times$2000 were collected from gastric mucosa and intestine. The dataset consists of 104 images with signet ring cells (positive) and 583 negative images, and there were 14,589 signet ring cells in total that were annotated with bounding boxes. The second task was colonoscopy tissue segmentation, where the dataset consists of 872 tissue sub-images from 476 patients, with an average size of 5000$\times$5000. The challenge received 234 effective submissions from 32 participating teams, where top-performing teams developed advancing approaches and tools for the CAD of digestive pathology.

\subsection{The StructSeg Challenge}
  
  The StructSeg 2019 Challenge\footnote{https://structseg2019.grand-challenge.org/} aimed to set up tasks for evaluating algorithms for automatic segmentation of gross target volume (GTV) and organs-at-risk (OAR) of nasopharynx cancer and lung cancer, for radiation therapy planning. For nasopharynx cancer, head and neck CT scans with 22 annotated OARs were provided, and for lung cancer, chest CT scans with 6 annotated OARs were provided. Both datasets have 50 scans for training and 10 scans for testing, respectively. The challenge attracted 574 teams for registration, and 45 teams submitted valid docker containers for evaluation. % For the OAR segmentation of head and neck region, the best performing team achieved an average Dice of 0.8109.

\section{Clinical Applications}~\label{sec:clinical_app}
The most important high-level goal of SenseCare is to serve as a platform for clinical research in various applications. In this section, we give several examples of user scenarios where SenseCare is adapted for various  clinical applications.
	
	\iffalse
	\textbf{Radiology Image}
	
	\begin{figure}[!htb]
		\begin{minipage}[b]{1.0\linewidth}
			\centering
			\centerline{\epsfig{figure=imageplatform.png,width=8.5cm}}
		\end{minipage}
		\caption{SenseCare 3D Medical Image Post-Processing}
	\end{figure}
	\fi
	
	\iffalse
	\textbf{Pathological Image}
	
	\begin{figure}[!htb]
		\begin{minipage}[b]{1.0\linewidth}
			\centering
			\centerline{\epsfig{figure=pictures/pathologyplatform.png,width=8.5cm}}
		\end{minipage}
		\caption{SenseCare Pathological Image}
	\end{figure}
	\fi
\subsection{Lung-oriented Application}
\begin{figure}
    \begin{minipage}[b]{1.0\linewidth}
        \centering
        \centerline{\epsfig{figure=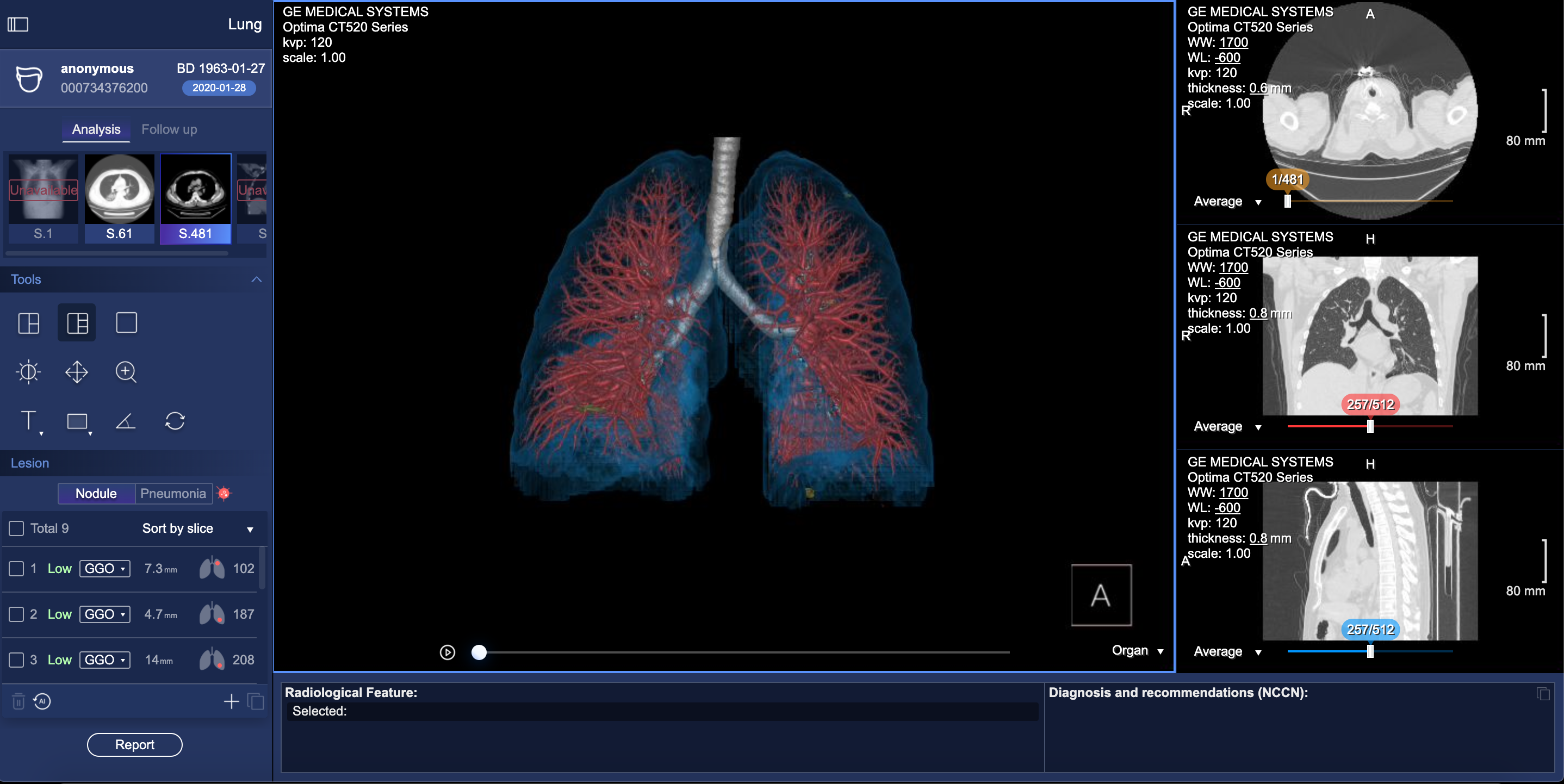,width=\linewidth}}
    \end{minipage}
    \caption{Intelligent analysis of lung nodules and preoperative planning in SenseCare.}
    \label{fig:lung}
\end{figure}
	Lung cancer is a malignant lung tumor and becomes the leading cause of death in China. It has a five-year survival rate of 16\%-17\%, which is the lowest of all types of cancers. Researches have shown that early diagnosis and intervention will improve the five-year survival rate up to 54\%. Although computer-assisted image analysis have been adopted to facilitate the lung cancer detection and diagnosis in an early stage, the various sizes, shapes and types of pulmonary nodules are placing great burdens on clinical physicians, thus causing misdiagnosis and missed diagnosis due to their fatigue and overwork. For lung cancer treatment, current clinic practice has a lack of effective preoperative planning tools. As a result, respiratory physicians find it difficult to personalize treatment plans in the surgical practices. 
	\\
	\\
	Based on its leading algorithms, SenseCare supports a thorough research and analysis of pulmonary nodules and lesions~\cite{wang2020tmi} by automatic detection, segmentation, and quantitative analysis. It can automatically detect and locate the nodules~\cite{Yu2019a,Luo2022mia} and then provide further quantitative information of each nodule such as its volume and density, in addition to qualitative estimation of its type and malignancy. As shown in Fig.~\ref{fig:lung_nodules}, four kinds of nodules are automatically detected and distinguished by SenseCare~\cite{Zhang2019}. In addition, it offers intelligent three-dimensional surgical planning for clinicians. By providing 3D visualization of thoracic anatomies including nodules, respiratory tracts  and pulmonary vessels, as well as bronchoscopy intervention path recommendation, SenseCare facilitates clinicians performing preoperative planning in a more intuitive and convenient way. Fig.~\ref{fig:lung} displays the user interface of a lung image analysis and preoperative planning application in SenseCare.

	\subsection{Pathology-related Application}
	
	%\begin{figure}[htbp]
	%\begin{minipage}[b]{1.0\linewidth}
	%  \centering
	% \centerline{\epsfig{figure=pictures/pathology.png,width=\linewidth}}
	%\end{minipage}
	%\caption{Pathology}
	%\label{pathology}
	%\end{figure}
	%image retrieval~\cite{Zhang2015c}. 
	
\begin{figure}
		\begin{minipage}[b]{1.0\linewidth}
			\centering
			\centerline{\epsfig{figure=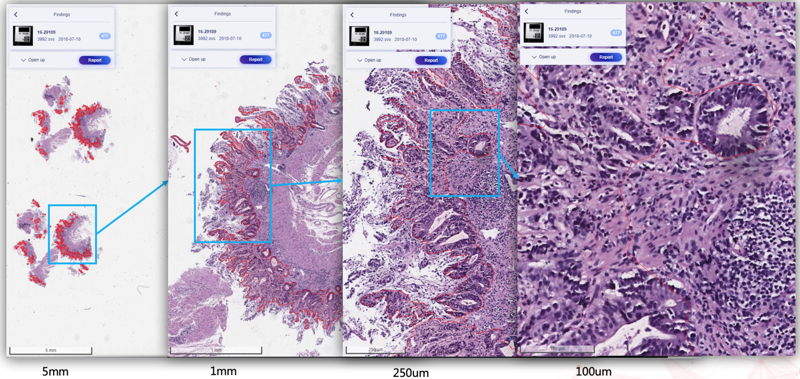,width=\linewidth}}
		\end{minipage}
		\caption{Pathological image analysis in SenseCare.}
		\label{fig:pathology_app}
	\end{figure}

	Pathological diagnosis is regarded as the most reliable criteria for diagnosis of cancers such as Gastrointestinal cancer. An early diagnosis is critical for good prognosis. However, it is labor-insensitive and time-consuming to manually detect lesion areas and pathological cells from 100,000$\times$100,000-pixel whole slide pathological images, which easily leads to fatigue of human and cannot satisfy the increasing demand for pathological diagnosis due to a lack of well trained experts in developing countries. 
	%Most of current workstations focus only on radiology image analysis. In addition to support of DICOM images, SenseCare has incorporated pathological analysis. In this part we will introduce a typical user scenario concerning digital pathological slides. 
	%\\
	%\\Gastrointestinal cancer, malignant conditions of the gastrointestinal tract and accessory organs of digestion, constitutes a major cancer type in China, whose incidence rates ranks second place and mortality rate ranks fourth. Actually, if detected in an early stage, most of the gastrointestinal cancers can be cured or lightened. However, given the shortage of pathologists and the limited resources available, few patients have the chance to schedule a biopsy. 
	%\\
	%\\Although pathological diagnosis is regarded as the most reliable method for diagnosis, detecting lesion areas and pathological cells from 100,000 x 100,000-pixel pathological images poses challenges and leads to doctors’ fatigue, resulting in misdiagnosis and missed diagnosis. On the other hand, the cultivation of pathological professionals is on the long run. 
	\\
	\\SenseCare-Pathology is designed to support pathological diagnosis covering gastrointestinal and cervical diseases at both tissue and cell levels~\cite{Li2021miccai,Qu2021npg}. It enables pathology department of hospitals and the third-party pathological diagnosis centers to perform efficient and intelligent analysis of pathological images by providing AI-based functional modules such as lesion and cell detection~\cite{Li2019a}, nuclei/gland segmentation~\cite{Qu2019,Qu2020tmi} and cell segmentation~\cite{Zhang2015c}, etc. Especially, SenseCare supports efficient retrieval of pathological images from a large-scale database~\cite{Zhang2015c,Li2018b, Zhang2015d}. Users can employ the retrieval module to search for similar images for a given input image. These modules help clinical researchers build capability to conduct large-scale cancer screening projects. Fig.~\ref{fig:pathology_app} shows an example of lesion localization from pathological images. 
	
	%\begin{figure}[htbp]
	%\begin{minipage}[b]{1.0\linewidth}
	%  \centering
	% \centerline{\epsfig{figure=pictures/specimen2.png,width=\linewidth}}
	%\end{minipage}
	%\caption{specimen2}
	%\label{specimen2}
	%\end{figure}

	\subsection{Pelvic Tumor Surgical Planning}
	\begin{figure}
		\begin{minipage}[b]{1.0\linewidth}
			\centering
			\centerline{\epsfig{figure=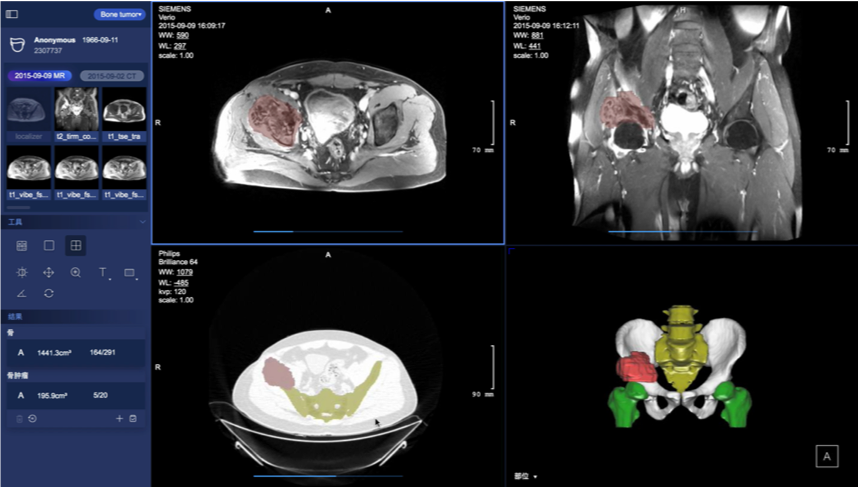,width=\linewidth}}
		\end{minipage}
		\caption{Pelvic tumor surgical planning system in SenseCare.}
		\label{fig:pelvic_tumor}
	\end{figure}
	Pelvic tumor is one of highly malignant tumors, and surgical treatment is the most effective treatment for pelvis tumors, where an accurate preoperative simulation and planning based on segmentation and modeling of the tumor is critical for the success of surgery. 
	%Many patients with bone tumors have to undergo amputation. Even those who can be treated with surgery, their mobility might also be reduced. Now an increasing number of doctors use three-dimensional models to conduct preoperative simulations. However, three-dimensional printing planning involves radiology, orthopedics, and engineers, thus negatively affecting the efficiency and effects of the planning process.
	%\\ 
	%\\Traditionally, there are five steps involving three-dimensional models generation. At first, various important parameters should be measured from X-ray images. Then orthopedics specialists are supposed to delineate the bone tumor areas from MR images, and then segregate and distinguish different bones from CT images. Registration between multimodal images, CT-MR data registration will be accomplished afterwards. Based on these results, three-dimensional engineers can reconstruct high-resolution data and generate three-dimensional printing models. Those cumbersome and tedious procedures involves multi-party cooperation and inevitably bring high communication cost. Too much time in preoperative planning may lead to deterioration of health conditions and previous plan could not be suitable for the patient anymore.~\cite{Ronneberger2015}
	\\
	\\
	SenseCare provides an intelligent preoperative surgical planning for  for limb salvage surgery of malignant pelvic tumors, where the accuracy and efficiency is improved by  our deep learning-based algorithms. The surgical planning workflow mainly consists of three parts. First, the pelvic tumor is segmented from MR scans with a U-Net like model~\cite{Ronneberger2015}. Then the pelvic bone is segmented from CT scans with a CNN combined with self-attention blocks~\cite{Fu2018}. Finally, a robust rigid/affine inverse-consistency registration method that is an extension of SymMirorr~\cite{Rivest-Henault2015} is conducted to align the corresponding MR-CT pair. Based on the registered CT-MR pair and the corresponding segmentation results, surgeons and radiologists could make accurate preoperative surgical planning rapidly. Fig.~\ref{fig:pelvic_tumor} shows the user interface of pelvic tumor surgical planning system in SenseCare. The system efficiently cuts down the multi-party communication costs between radiology, orthopedics and 3D printing centers, and ultimately reduces doctor’s workload and facilitates patient-tailored treatment.  It took only 15 minutes to complete the surgical planning for pelvic tumor resection, which is a dramatic acceleration compared with the 2-day time span in a traditional workflow~\cite{Qu2021MIA}.

	%SenseCare-Bone Tumor provides “one-click solution” for bone tumor surgical planning. The system automatically analyzes multimodality images and executes surgical planning steps including image registration, fusion, segmentation and 3D model generation within 1 minute. 

\subsection{Automatic Diagnosis of Coronary Artery Disease}
\begin{figure}
    \begin{minipage}[b]{1.0\linewidth}
        \centering
        \centerline{\epsfig{figure=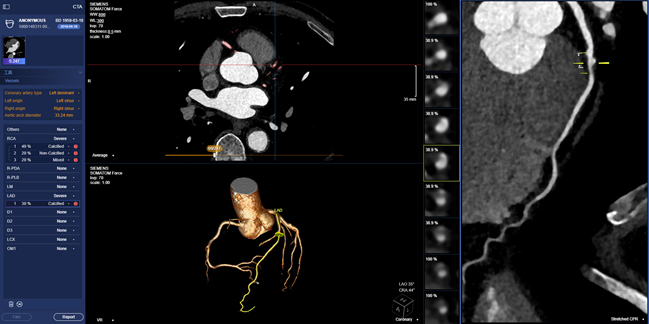,width=\linewidth}}
    \end{minipage}
    \caption{Coronary artery application in SenseCare.}
    \label{fig:coronary_arteries}
\end{figure}
Cardiovascular diseases (CVDs) are the most common causes of death throughout the world. Non-invasive morphological and functional assessment of cardiovascular structures plays an important role in diagnosis and treatment of CVDs. However, accurate and robust reconstruction of cardiovascular system  from Computed Tomography Angiography (CTA) is laborious and time-consuming due to the large variation of heart and low contrast between arteries and other tissues. SenseCare provides a full-stack solution towards automatic diagnosis of coronary artery disease, including fully automatic segmentation of 3D whole heart and coronary arteries, extraction of coronary artery centerlines, labeling of important artery branches, reconstruction of MPR and CPR, real-time volumetric rendering, detection of plaques and quantification of stenosis, and automatic generation of diagnose reports.
	\\
	\\More specifically, we propose a novel cascaded method for automatic segmentation of 3D whole heart~\cite{Wang2021heart} and coronary arteries, which is integrated into a semi-supervised framework~\cite{Tarvainen2017} and has achieved state-of-the-art performance using very few manual annotations. A unified deep reinforcement learning~\cite{VanHasselt2016} framework is proposed to automatically traverse tree-structure centerlines of coronary arteries.  Pixel-level segmentation followed by 3D classification and segmentation of point sets~\cite{Qi2017} is proposed for coronary artery labeling. %We adopt a GPU-accelerated framework for efficient reconstruction of MPR/CPR and real-time volumetric rendering of the segmented cardiovascular structures, which allows dozens of doctors to use in the same time. 
	We utilize a recurrent CNN to automatic detect and classify the type of coronary artery plaque~\cite{Zreik2019}, and the degree of coronary artery stenosis is quantified by a multi-class segmentation of plaque and vessel lumen from the reconstructed probe images. Finally, a structured diagnose report is summarized based on the results obtained from all steps mentioned above, and the whole procedure is finished within one minute.  Fig.~\ref{fig:coronary_arteries} shows a snapshot of the coronary artery application in SenseCare. 
	
\subsection{Lower Limb-oriented Application}
\begin{figure}
    \begin{minipage}[b]{1.0\linewidth}
        \centering
        \centerline{\epsfig{figure=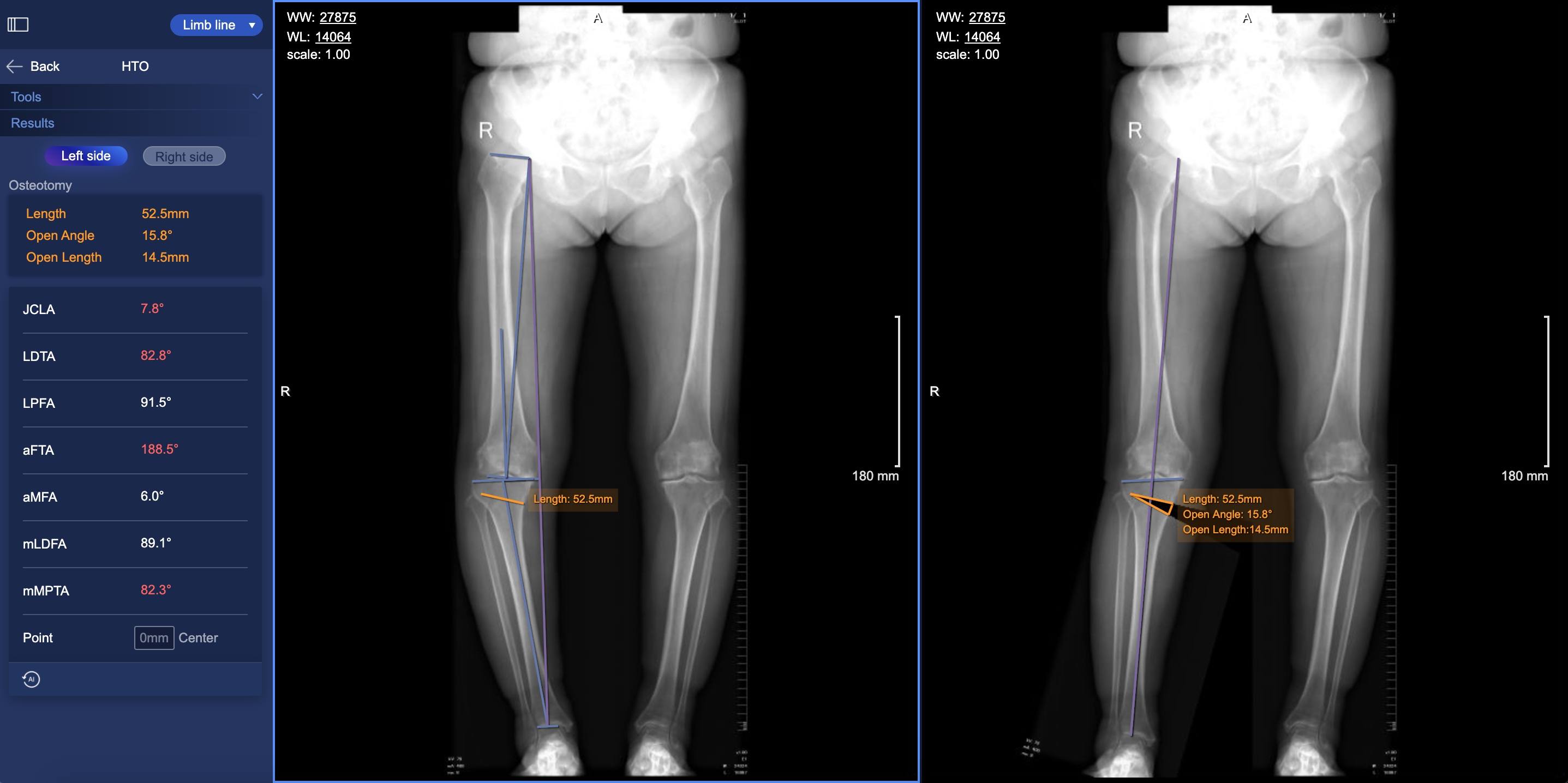,width=\linewidth}}
    \end{minipage}
    \caption{Lower limb application in SenseCare. (a) shows limb parameters measured on (b). (b) and (c) show the detected landmarks and HTO planning, respectively.}
    \label{fig:lower_limb}
\end{figure}

\begin{figure}
    \begin{minipage}[b]{1.0\linewidth}
        \centering
\centerline{\epsfig{figure=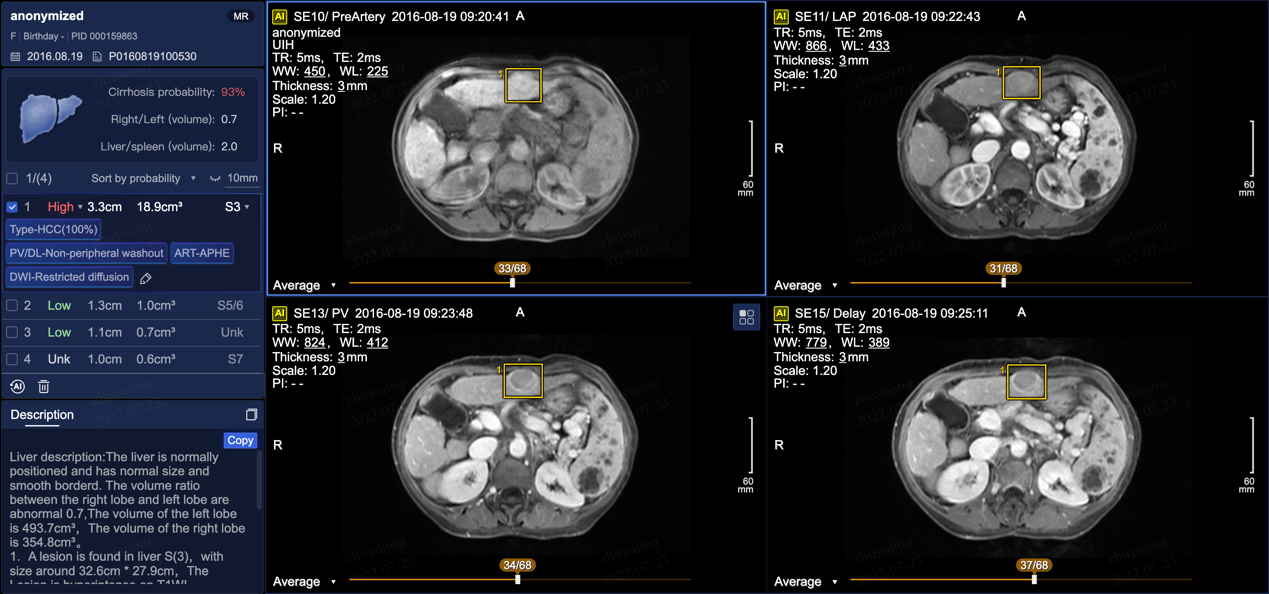,width=\linewidth}}
    \end{minipage}
    \caption{Intelligent diagnosis system of liver cancer in SenseCare.}
    \label{fig:liver_tumor}
\end{figure}
Knee Osteoarthritis (OA) is an important public health issue that causes chronic disability. Lower limb osteotomies is a well-established and commonly utilized technique in medial knee osteoarthritis secondary to different forms of knee joint malalignments, trying to establish a better alignment by passing the load bearing leg axis through hip, knee and ankle joint. The measurement of lower limb axial alignment with radiographs is a critical step in the preoperative planning to exactly define the characteristic of the osteotomy.
\\
\\SenseCare's lower limb surgical planning system provides advanced AI algorithms for automatic radiograph analysis such as segmentation of femur and tibia  and detection of lower limb landmarks for accurate measurement of the critical angles and distances of lower limbs. The results are used for planning of surgical treatments such as High Tibial Osteotomy (HTO), as shown in Fig.~\ref{fig:lower_limb}. The segmentation of femur and tibia is achieved by a U-Net~\cite{Ronneberger2015} followed by post-processing based on connected components to discriminate the left and right instances. Localization of landmarks is based on a coarse-to-fine recurrent network~\cite{Chen2017c} combined with Gaussian heatmap regression~\cite{ONeil2019}. The heatmap corresponding to a landmark location is a sum of multiple Gaussian functions centered at that landmark. At inference time, the largest values in the heatmap are taken as the  detected landmark positions. %Besides, the coarse and fine reliability are the similarity metrics between the predicted heatmaps and the ideal heatmaps target for the predicted landmark positions. Finally, when both the coarse and fine reliability are higher than the threshold, the fine landmark position will be chosen as the final prediction, otherwise the coast landmark position will be chosen. Consequently, 
Such a strategy helps to achieve stable and accurate limb landmark detection results that ensure the reliability of downstream surgical planning.

\begin{figure}
    \begin{minipage}[b]{1.0\linewidth}
        \centering
\centerline{\epsfig{figure=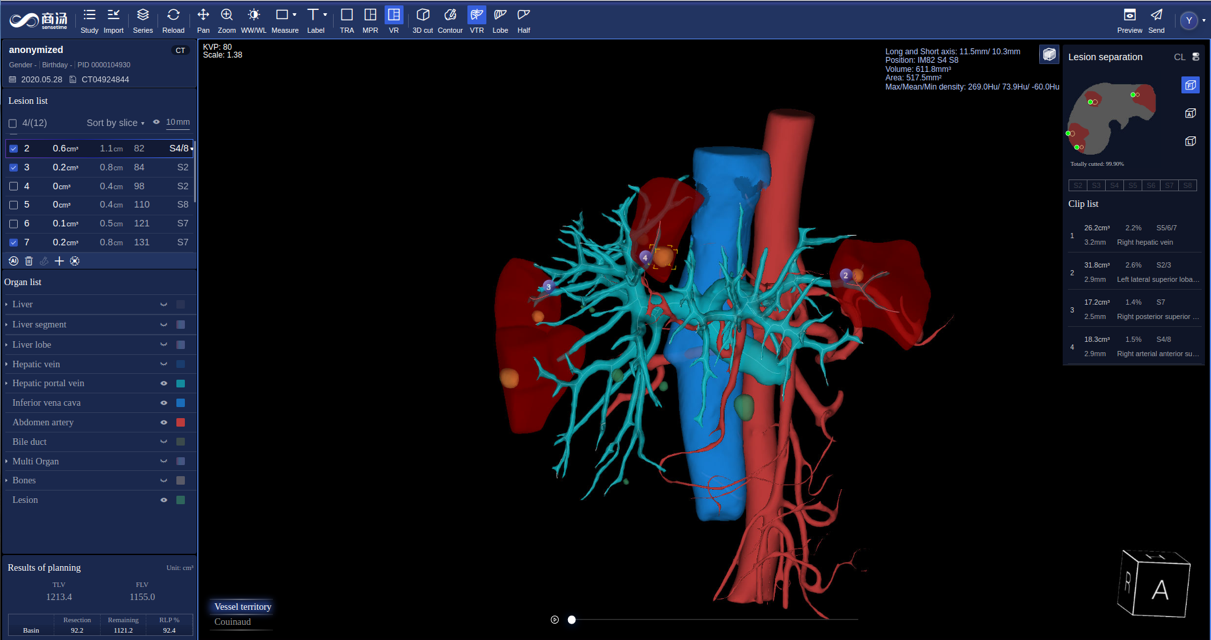,width=\linewidth}}
    \end{minipage}
    \caption{Liver surgical planning system in SenseCare.}
    \label{fig:liver_vessel}
\end{figure}
\subsection{Automatic Diagnosis of Liver Cancer}
Primary liver cancer, mainly hepatocellular carcinoma (HCC), is the sixth most prevalent malignancy and third leading cause of cancer-related death worldwide. The prognosis is poor owing to the high recurrence in 60\%-70\% of patients within 5 years after curative surgery~\cite{Wang2022jhc}. Diagnosis of liver cancer is important for treatment decision.  For accurate diagnosis, radiologists need to handle multiple CT or MRI sequences to analyze characteristics of liver tumors. 
\\
\\SenseCare provides a closed-loop platform for intelligent diagnosis of liver cancer using multi-phase dynamic CT scans and multi-sequence MRI images. As shown in Fig.~\ref{fig:liver_tumor}, it has an integrated pipeline for automatic sequence alignment, detection and localization of tumors~\cite{Wang2021ar}. Moreover, it provides morphology and density measurements for assessment of cancer type, malignancy and LI-RADS symptom~\cite{Wang2021ii}. It also implements automatic segmentation and volume measurements of abdomen organs, and risk assessment of cirrhosis and fatty liver. Finally, a structured and standardized diagnostic report is generated. Therefore, SenseCare serves as a promising tool to help radiologists locate and identify the suspected liver tumors more efficiently and effectively.

\subsection{Liver Surgical Planning}
Surgical treatment is one of the most important treatment methods for liver cancer. However, the liver's complex structure brings great challenges for ablation and resection treatments of liver tumor. 3D visualization of tumors and adjacent blood vessels has been included in the first-level recommendation of expert consensus on precision therapy of liver cancer. However, existing surgical planning products do not achieve automatic and accurate segmentation, which require a lot of manual interaction and modification.

\begin{figure}
    \begin{minipage}[b]{1.0\linewidth}
        \centering
\centerline{\epsfig{figure=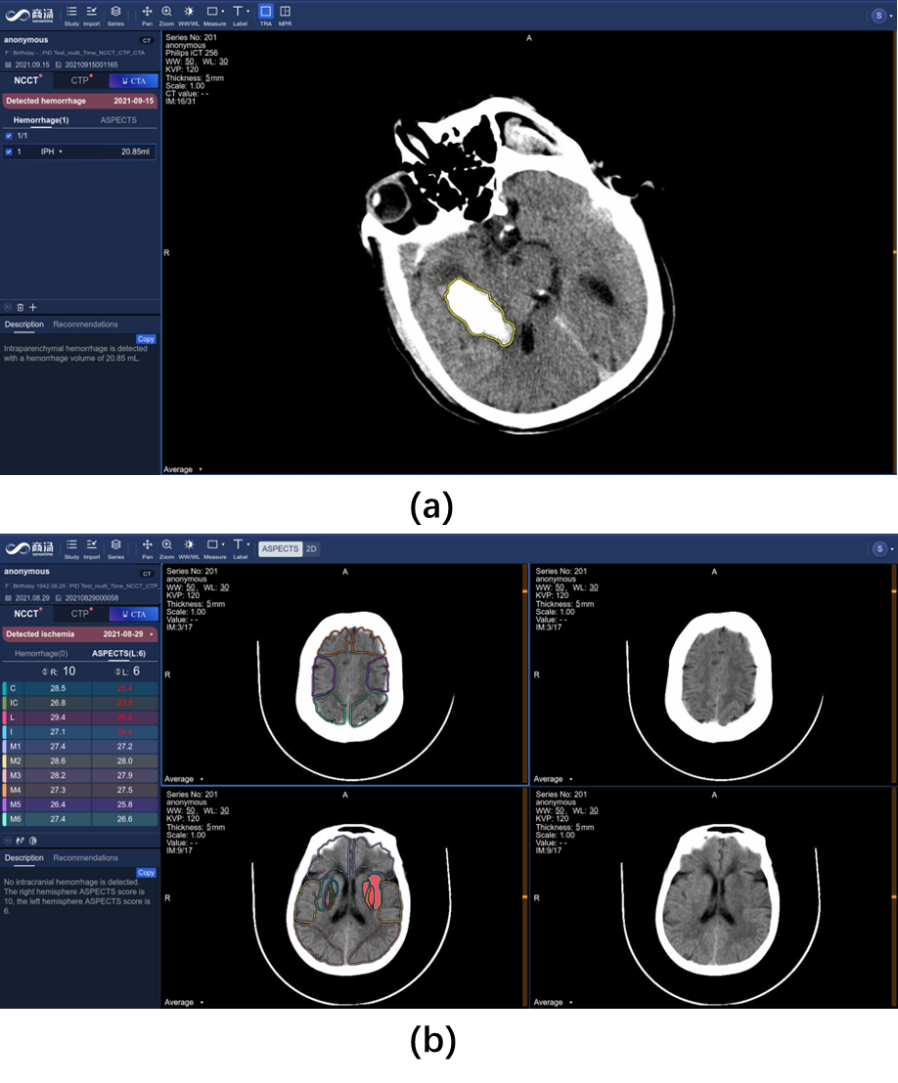,width=\linewidth}}
    \end{minipage}
    \caption{Stroke diagnosis in SenseCare. (a) hemorrhage diagnosis. (b) ASPECT score measurement for ischemic stroke.}
    \label{fig:stroke_1}
\end{figure}

SenseCare provides a quantitative 3D modeling of the liver for highly automated liver-specific analysis, which is shown in Fig.~\ref{fig:liver_vessel}. It helps hepatobiliary surgeons perform interactive analysis and treatment planning. In detail, it supports efficient automatic segmentation and quantitation of tumors, vessels, ducts and abdomen organs, and interactive refinement of segmentation results. It supports pre-surgical planning for liver resection, such as vessel territory resection, tri-lobe resection, hemi hepatectomy. The system also aides in precise needle placement relative to tumors and surrounding tissues for ablation treatment. It helps clinicians get a whole picture of the abdomen in a quicker and more comprehensive paradigm, which improves the safety of operations and ensures the treatment outcome.

\subsection{Stroke Diagnosis}
Stroke is a severe cerebrovascular disease globally, with high incidence, high  disability rate and mortality~\cite{Kissela2012}. %Stroke includes hemorrhagic stroke and ischemic stroke, and 
Ischemic stroke is the most common type of stroke, accounting for 75-85\% of all stroke cases. %Ischemic stroke is a disorder of blood supply to the brain that results in tissue hypoxia (hypoperfusion) and tissue death within a few hours. %Stroke treatment is divided into treatment within the time window (0 to 4.5 h) and treatment beyond the time window (4.5 to 24h).
Early diagnosis and treatment are crucial for the rehabilitation of stroke patients, and the detection and quantitative evaluation of stroke lesions by medical imaging is of great significance for accurate diagnosis and treatment decision.
\\

\begin{figure}
    \begin{minipage}[b]{1.0\linewidth}
        \centering
\centerline{\epsfig{figure=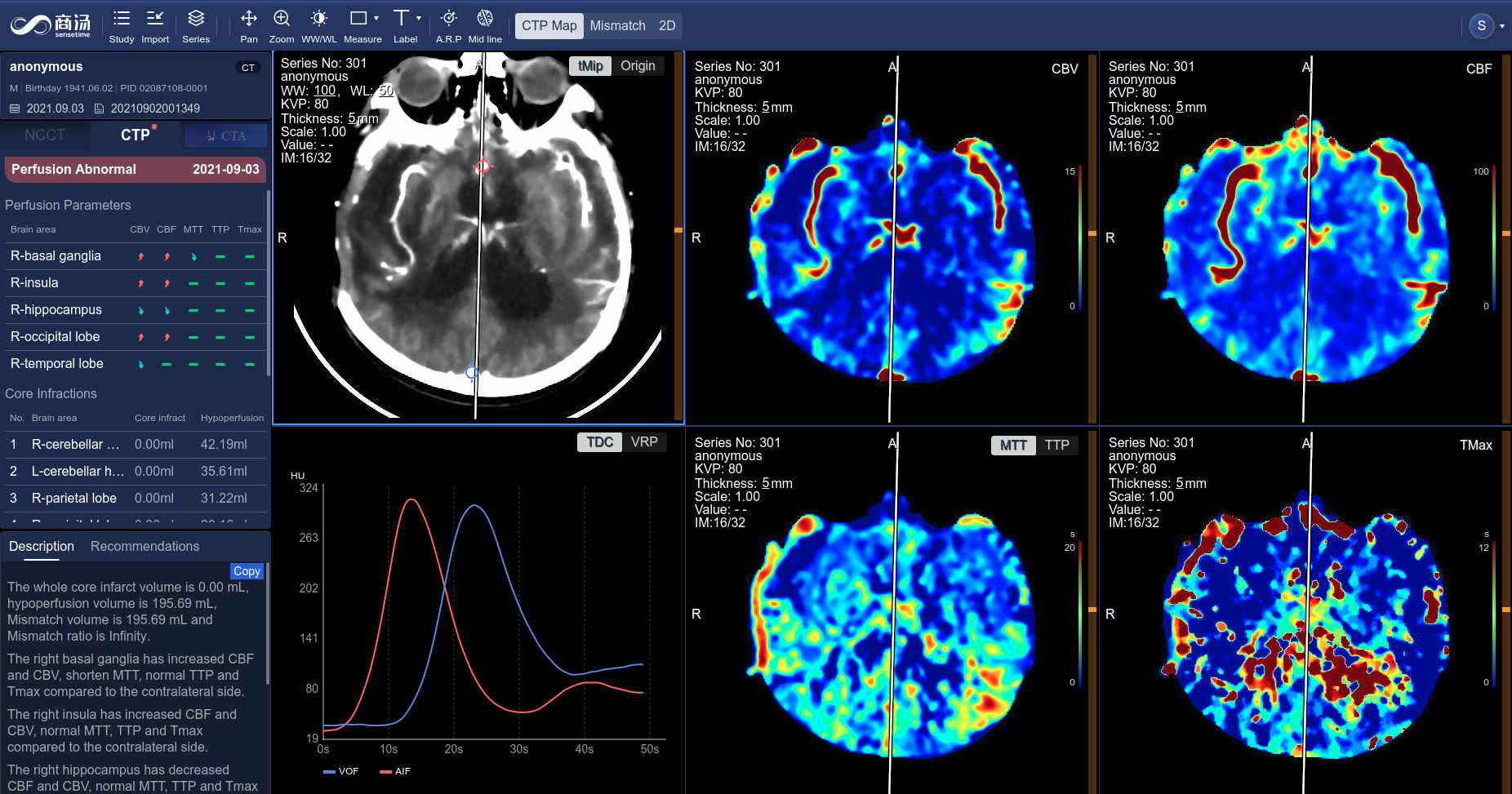,width=\linewidth}}
    \end{minipage}
    \caption{CT perfusion analysis system of ischemic stroke in SenseCare.}
    \label{fig:stroke_2}
\end{figure}
\begin{figure}
    \begin{minipage}[b]{1.0\linewidth}
        \centering
\centerline{\epsfig{figure=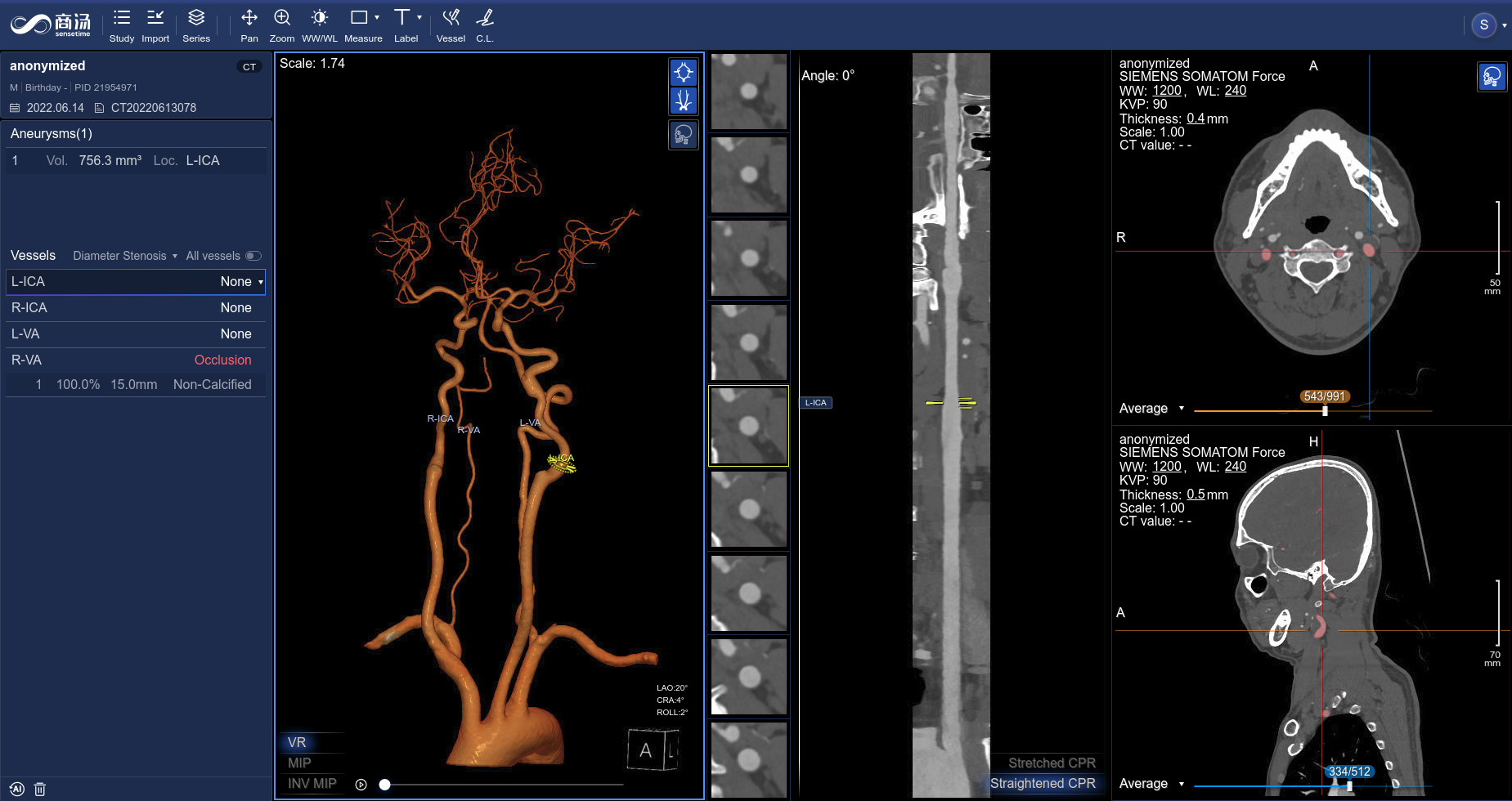,width=\linewidth}}
    \end{minipage}
    \caption{Head and neck CTA analysis system in SenseCare.}
    \label{fig:stroke_3}
\end{figure}

The SenseCare platform provides a one-stop stroke diagnosis solution based on different types of CT imaging. First, it provides automatic analysis of intracerebral hemorrhage and ischemic stroke using non-contrast CT. As shown in Fig.~\ref{fig:stroke_1}(a), for intracerebral hemorrhage, it can detect the hemorrhage and calculate the volume of bleeding automatically, and classify the hemorrhage into five types.
For ischemic stroke, the ASPECT score is automatically measured for preliminary assessment of the severity of ischemia, as shown in Fig.~\ref{fig:stroke_1}(b). 
Second, for ischemic stroke, the CT perfusion analysis system can decode raw 4D CT perfusion images into perfusion parameter maps (CBV, CB, MTT and TMAX) for accurate diagnosis, as shown in Fig.~\ref{fig:stroke_2}. It can quantify the volume~\cite{Wang20201stroke} and mismatch between the infarct core and the ischemic penumbra. Thirdly, for thrombectomy patients, SenseCare  features a head and neck CTA reconstruction and analysis system~\cite{Fu20201stroke} for rapid vascular reconstruction and localization of plaque occluded areas and aneurysm, as shown in Fig.~\ref{fig:stroke_3}. The holistic solution can be applied to stroke centers to improve their treatment capacity.

\subsection{Radiotherapy Contouring}
\begin{figure}
	\begin{minipage}[b]{1.0\linewidth}
		\centering
		\centerline{\epsfig{figure=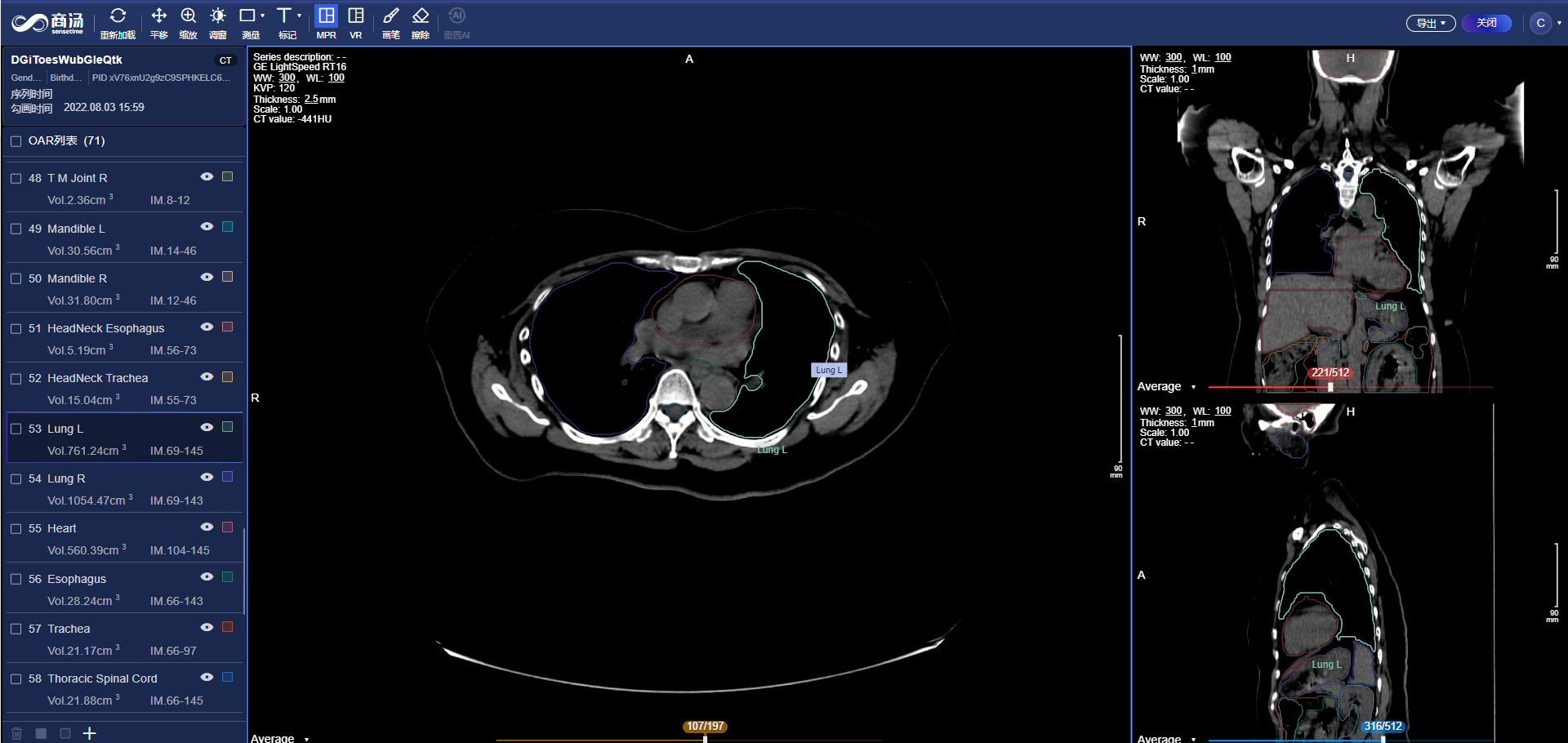,width=\linewidth}}
	\end{minipage}
	\caption{Radiotherapy contouring system in SenseCare.}
	\label{fig:rt}
\end{figure}
Radiotherapy is an important and widely-adopted treatment for cancers.  In radiotherapy treatment planning, one of the most critical step is the delineation of the organs-at-risk (OAR) and target volumes, the quality of which is a core factor affecting the efficacy and side effects of radiotherapy. Clinically, radiologists have to spend several hours for manual delineation, due to the large number of OARs, complex shapes of cancers and large size of 3D volumes. As a result, it requires great time and efforts, as well as a high level of professionalism for the radiologist.

SenseCare Radiotherapy Contouring system aids in the radiologists by automatically contouring OARs across the whole body, including the head and neck \cite{fu2022hmrnet,gao2021focusnetv2,lei2021automatic,liu2020csaf}, chest \cite{liu2021automatic} and abdomen \cite{huang2020multi}, as well as common targets including breast cancer, rectal cancer \cite{yang2022automatic}, etc.  The system can be seamlessly connected to different kinds of CT machines and treatment planning systems (TPS), automatically run AI calculation in background, support a 3D interactive view and editing the delineated structures, and export the results back via the standard RT structure DICOM protocol, in a complete closed loop of workflow.

\section{Conclusion and Future Work}\label{sec:conclusion}
With the development of medical imaging techniques and artificial intelligence, AI-based medical image computing systems have an increasing potential for more intelligent diagnosis and treatment in clinic practice. However, existing deep learning platforms rarely support specific clinical applications while most current medical image analysis platforms are not well supported by advanced AI algorithms. To boost clinical research towards smart healthcare, SenseCare is developed as a generic research platform for intelligent medical image computing  that can support various research needs across different medical disciplines. 
\\
\\SenseCare provides a large set of AI algorithms for medical image segmentation, registration, detection and other tasks that can help clinicians and radiologists conduct various clinic-oriented translational research programs, such as lung cancer diagnosis and surgical planning, efficient pathological image analysis, pelvic tumor and limb surgical planning, coronary artery disease diagnosis and modeling, etc. In addition to the built-in AI algorithms, SenseCare also provides several tools for users to develop and deploy customized AI models efficiently.  The AI toolkits and other appearing functional modules such as advanced visualization, web-based access and multi-center deployment in SenseCare can efficiently boost clinical research programs and applications towards smart healthcare. 
\\
\\In the future, we will continue working on improving the reliability and stability of SenseCare while enlarging its capability and functionality for wider use scenarios.
It should be emphasized that SenseCare is temporarily only used for research purposes and we are conducting clinical studies under regulations, to translate AI techniques into a revolution in clinic practice with more efficient,  accurate and intelligent healthcare.
	
\section{Acknowledgments}
We would like to express our gratitude to our research and product team and clinical collaborators, without whom we cannot develop our algorithms and user-interface so efficiently and rapidly. 
	
% -------------------------------------------------------------------------
\bibliographystyle{IEEEbib}
\bibliography{sense_care}
%\bibliography{D:/Documents/texstudio/latex_reference/sense_care}
\end{document}